

\magnification 1200
%

%
\font\eightrm=cmr8
\font\eighti=cmmi8
\font\eightsy=cmsy8
\font\eightbf=cmbx8
\font\eighttt=cmtt8
\font\eightit=cmti8
\font\eightsl=cmsl8
\font\sixrm=cmr6
\font\sixi=cmmi6
\font\sixsy=cmsy6
\font\sixbf=cmbx6
\catcode`@11
\newskip\ttglue
\font\grrm=cmbx10 scaled 1200

\def\eightpoint{\def\rm{\fam0\eightrm}
\textfont0=\eightrm \scriptfont0=\sixrm \scriptscriptfont0=\fiverm
\textfont1=\eighti \scriptfont1=\sixi \scriptscriptfont1=\fivei
\textfont2=\eightsy \scriptfont2=\sixsy \scriptscriptfont2=\fivesy
\textfont3=\tenex \scriptfont3=\tenex \scriptscriptfont3=\tenex
\textfont\itfam=\eightit \def\it{\fam\itfam\eightit}
\textfont\slfam=\eightsl \def\sl{\fam\slfam\eightsl}
\textfont\ttfam=\eighttt \def\tt{\fam\ttfam\eighttt}
\textfont\bffam=\eightbf
\scriptfont\bffam=\sixbf
\scriptscriptfont\bffam=\fivebf \def\bf{\fam\bffam\eightbf}
\tt \ttglue=.5em plus.25em minus.15em
\normalbaselineskip=6pt
\setbox\strutbox=\hbox{\vrule height7pt width0pt depth2pt}
\let\sc=\sixrm \let\big=\eightbig \normalbaselines\rm}
\newinsert\footins
\def\newfoot#1{\let\@sf\empty
  \ifhmode\edef\@sf{\spacefactor\the\spacefactor}\fi
  #1\@sf\vfootnote{#1}}
\def\vfootnote#1{\insert\footins\bgroup\eightpoint
  \interlinepenalty\interfootnotelinepenalty
  \splittopskip\ht\strutbox 
  \splitmaxdepth\dp\strutbox \floatingpenalty\@MM
  \leftskip\z@skip \rightskip\z@skip
  \textindent{#1}\footstrut\futurelet\next\fo@t}
\def\fo@t{\ifcat\bgroup\noexpand\next \let\next\f@@t
  \else\let\next\f@t\fi \next}
\def\f@@t{\bgroup\aftergroup\@foot\let\next}
\def\f@t#1{#1\@foot}
\def\@foot{\strut\egroup}
\def\footstrut{\vbox to\splittopskip{}}
\skip\footins=\bigskipamount 
\count\footins=1000 
\dimen\footins=8in 

\def\ref#1{$^{#1}$}
\def\flex{\raise 6pt\hbox{$\leftrightarrow $}\! \! \! \! \! \! }
\def\oversome#1{ \raise 8pt\hbox{$\scriptscriptstyle #1$}\! \! \! \! \! \! }

\newbox\bigstrutbox
\setbox\bigstrutbox=\hbox{\vrule height10pt depth5pt width0pt}
\def\bigstrut{\relax\ifmmode\copy\bigstrutbox\else\unhcopy\bigstrutbox\fi}
\def\refer[#1/#2]{ \item{#1} {{#2}} }
\def\rev<#1/#2/#3/#4>{{\it #1\/} {\bf#2}, {#3}({#4})}
\def\boxit#1{\vbox{\hrule\hbox{\vrule\kern3pt
\vbox{\kern3pt#1\kern3pt}\kern3pt\vrule}\hrule}}

\def\2figure#1#2#3#4{\vbox{ \hrule width#1truecm \hbox{\vrule height#2truecm
\hskip #1truecm
\vrule height#2truecm }\hrule width#1truecm \hbox{\vrule\vbox{\hsize #1truecm
\baselineskip=10pt
\noindent\strut#3}\vrule}\hrule width#1truecm
\hbox{\vrule\vbox{\hsize #1truecm
\baselineskip=10pt
\noindent\strut#4}\vrule}\hrule width#1truecm  }}
\def\3figure#1#2#3#4#5{\vbox{ \hrule width#1truecm \hbox{\vrule height#2truecm
\hskip #1truecm
\vrule height#2truecm }\hrule width#1truecm \hbox{\vrule\vbox{\hsize #1truecm
\baselineskip=10pt
\noindent\strut#3}\vrule}\hrule width#1truecm
 \hbox{\vrule\vbox{\hsize #1truecm
\baselineskip=10pt
\noindent\strut#4}\vrule}
\hrule width#1truecm \hbox{\vrule\vbox{\hsize #1truecm
\baselineskip=10pt
\noindent\strut#5}\vrule}\hrule width#1truecm  }}

\def\sqr#1#2{{\vcenter{\hrule height.#2pt
   \hbox{\vrule width.#2pt height#1pt \kern#1pt
    \vrule width.#2pt}
    \hrule height.#2pt}}}


\def\smin{\,\raise 0.06em \hbox{${\scriptstyle \in}$}\,}
\def\smsubset{\,\raise 0.06em \hbox{${\scriptstyle \subset}$}\,}

\def\Natural{\hbox{\hskip 1.5pt\hbox to 0pt{\hskip -2pt I\hss}N}}

\def\Rational{\hbox{\hbox to 0pt{\hskip 2.7pt \vrule height 6.5pt
                                  depth -0.2pt width 0.8pt \hss}Q}}
\def\Real{\hbox{\hskip 1.5pt\hbox to 0pt{\hskip -2pt I\hss}R}}
\def\Complex{\hbox{\hbox to 0pt{\hskip 2.7pt \vrule height 6.5pt
                                  depth -0.2pt width 0.8pt \hss}C}}


\nopagenumbers
\centerline {\grrm Correlators in non-critical superstrings including }

\centerline{\grrm the spinor emission vertex.}
\vskip 1cm
\centerline {\bf D. Dalmazi\newfoot{\ref*}{Supported by CNPq} and
E. Abdalla,\newfoot{\ref\dagger}{Partially supported by CNPq}}
\vskip .2cm

\centerline {Instituto de F\'\i sica, Univ. S\~ao Paulo, CP 20516,
S\~ao Paulo, Brazil}
\vskip .2cm

\centerline {\bf Abstract}
\vskip .5cm
We discuss the structure of correlators involving the spinor emission vertex
in non critical $N=1$ superstring theory.  The technique used in the
computation is the zero mode integration to arrive at the integral
representation, and later an analysis of the pole structure of the integrals
which are thus obtained. Our analysis has been done primarily for the 5-point
functions. The result confirms previous expectations and prepares
ground for a comparison with computations using matrix models techniques.

\vskip.5cm

\hfill Universidade de S\~ao Paulo\quad

\hfill IFUSP-preprint-1029\phantom{Paulo}\quad

\hfill February 1993\phantom{Paulo}\quad

\vfill

\eject
\countdef\pageno=0 \pageno=1
\newtoks\footline \footline={\hss\tenrm\folio\hss}
\def\folio{\ifnum\pageno<0 \romannumeral-\pageno \else\number\pageno \fi}
\def\advancepageno{\ifnum\pageno<0 \global\advance\pageno by -1
\else\global\advance\pageno by 1 \fi}

The recent discovery of the double scaling limit\ref{1} in matrix models has
led to precise computations of higher genus correlation functions in non
critical strings propagating in $d\le 2$-dimensional target spaces. From the
point of view  of the theory in the continuum space, however, such calculations
have been carried out only for the sphere (genus zero). The computations have
been performed either by integrating the Liouville zero mode\ref{2,3} with
the subsequent evaluation of the corresponding integrals, or more recently by
use of the ground ring structure\ref{4}, or also by  quantum group
technique\ref{5}. The agreement of all such approaches (see [2,6] for
comparison) encourage us to generalize such correlators on the sphere to the
case of the superstring. By use of the zero mode technique this has been done
in [7,8,9] by some groups, at least
in the Neveu-Schwarz sector in the $N=1$ case. The $N=2$ non-critical
theory is discussed in [10]. Almost simultaneously, a supersymmetric
version of the one matrix model has appeared in the literature\ref{11}, which
describes super $(p,q)$ minimal models with $p=4$, $q=4m$, $m=1,2,\cdots$
coupled to two dimensional supergravity. An identification of scaling operators
and comparison analogous to [12] can be successfully carried\ref{13},
but for a more precise comparison in the Ramond sector
we have to calculate further correlators involving the spinor emission vertex,
thus generalizing the calculations of [3]. The aim of this letter is to
calculate some of such correlators missing in the literature.

The supergravity part of the superstring is described by an $N=1$ super
Liouville theory, defined by the action (see e.g. [14])

$$S_{SL}={1\over 4\pi}\int d^2zd^2\theta \hat E \left( {1\over 2} D_\alpha
\Phi_{SL} D^\alpha \Phi_{SL} -Q\hat Y\Phi_{SL}- 4i\mu
e^{\alpha_+\Phi_{SL}}\right) \eqno(1)$$
where $\hat E$ is the superdeterminant of the superzweibein $E_{ab}$, and
$\hat Y$  is the supercurvature, and $\Phi_{SL}$ is the super Liouville
superfield,  which defines the super world sheet dynamics. The matter sector
is given by an $N=1$ superfield $\Phi_M$, with the action
$$S_{M}={1\over 4\pi}\int d^2zd^2\theta \hat E \left( {1\over 2} D_\alpha
\Phi_{M} D^\alpha \Phi_{M} -Q\hat Y\Phi_{M}\right) \eqno(2)$$
The matter sector has central charge $\hat c_M=1-8\alpha_0^2$, and $Q$,
$\alpha_\pm$ are defined by
$$\eqalignno{Q&=2\left( 1+\alpha_0^2\right)^{1\over 2}, &(3a)\cr
\alpha_\pm &=-{1\over 2}Q\pm \vert\alpha_0\vert &(3b)\cr}$$

The particle content of the 2D-superstring consists of  a scalar (NS-sector)
and a spinor (R-sector) particle in the space time. Both are massless. The
Neveu-Schwarz vertex (or scalar emission vertex) is simply the
supersymmetric extension of a planar wave (the ``tachyon"), being defined
by
$$T_k =\int d^2zd^2\theta\hat E e^{ik\Phi_M+\beta\Phi_{SL}}=\int d^2z
e^{ikX+\beta\phi}\left(\beta\psi+ik\xi\right)\left(\beta\overline\psi+ik
\overline\xi\right) \eqno(4)$$
where $\psi$, $\xi$ are  partners of $\phi$, $X$, forming the fermionic
and bosonic components of $\Phi_{SL}$, $\Phi_M$. The dressing $\beta (k)$ is
given by
$$E=\beta (k) +{Q\over 2}=\vert k-\alpha_0\vert\; .\eqno(5) $$
It will be necessary for
our purposes to have also another form of (4) which corresponds to the gauge
fixed $(\theta =0)$ version of the NS-vertex, $T_k^{(-1)}$, given by
$$T_k^{(-1)}=\int d^2z e^{-\sigma+ikx+\beta\phi}\; ,\eqno(6)$$
where $\sigma$ is related to the original supersymmetry ghosts via
bosonization (see [15] for details),  and the (holomorphic part of the)
propagator is given by
$$\left\langle\sigma(z)\sigma(w)\right\rangle=-\ln (z-w).\eqno(7)$$
Note that the definition of $\beta$ in (5) is still the same since
$e^{-\sigma}$ has the same conformal weight of $d^2\theta$ (in order to check,
one can use $\Delta\left( e^{a\sigma}\right)=-a(a+2)/2$).

In the next step we turn to the spinor emission vertex, which is more
complicated. For future convenience, we first bosonize
the two fermionic (Majorana) components $\psi$ and $\xi$ into a free massless
boson $h$, following the usual rules for the Dirac fermions $\psi \pm\xi$ (we
omit cocycles)
$$\psi\pm i\xi =\sqrt 2e^{\mp ih}.\eqno(8)$$
The propagator $\langle h(z)h(w)\rangle =-\ln (z-w)$ gives rise to the usual
fermionic propagators
$$\eqalign{\left\langle\psi (z)\psi (w)\right\rangle &=\left\langle\xi (z)
\xi (w)\right\rangle ={1\over z-w}\cr
\left\langle\psi (z)\xi (w)\right\rangle &= 0\; ,\cr}$$
analogous results hold for the anti-holomorphic part.
Following [15] we notice that states in the the Neveu-Schwarz (resp. Ramond)
sector have periodic (antiperiodic) boundary conditions, and the spinor
emission vertex which will correspond later to the emission of a fermionic
particle in the two dimensional embedding space, must interchange such boundary
conditions. Therefore, the spinor emission vertex must include a field
$S_\epsilon$ ($\epsilon =\pm 1$) --called spin field-- which introduces a cut
at the point of the emission, such that after a rotation of $2\pi$ one gets a
minus sign, in other words,
$$\psi^\mu(z)S_\epsilon (w)\sim
{\left(\gamma^\mu\right)_\epsilon^{\,\beta}S_\beta (z)\over (z-w)^{1\over
2}}\eqno(9)$$
where $\gamma^\mu$ are the two dimensional gamma matrices and $\psi^0=\psi$,
$\psi^1=\xi$. The solution to the above OPE constraint is given by
$$S_\epsilon=e^{{i\over 2}\epsilon h}$$
Due to supersymmetry, we need also a spinor field corresponding to the ghost
$\sigma$,
$$\Sigma_{-{1\over 2}}=e^{-{1\over 2}\sigma}\, ,  $$
Therefore we have
$$V_{-{1\over 2}}(k,\epsilon )=\int d^2z e^{-{1\over 2}
\sigma +{i\over 2}\epsilon h+ikx+ \beta\phi} \eqno(10)$$
 It is not difficult to check that $S_\epsilon$, as well as
$V_{-{1\over 2}}(k,\epsilon )$ behave like a Weyl spinor under $SO(1,1)$
generators $\colon\psi^\mu\psi^\nu\colon$. Imposing BRST invariance of
$V_{-{1\over 2}}(k,\epsilon )$ we have two different dispersion relations for
the different spin components $\epsilon =\pm 1$, as expected for a Weyl spinor
in two dimensions\newfoot{\ref {1)}}{The
BRST invariance automatically guarantees
supersymmetry for $V_{-{1\over 2}}(k,\epsilon )$. Notice that this is not
explicit, as in the case of $T(k)$ (see [15]).}:
$$E=\beta +{Q\over 2}= -\epsilon (k-\alpha_0)\eqno(11)$$
We are now in position to compute mixed correlation functions:
$${\cal A}_{\cal N}^{(n,{\cal N}-n)}
(k_1\cdots k_n,\epsilon_1\cdots\epsilon_m)=\left\langle
\prod_{i=1}^m V_{-{1\over 2}}(k_i,\epsilon_i ) \prod_{j=m+1}^m
T(k_j)\right\rangle_\mu\; .\eqno(12)$$
After integrating over the matter ($X_0$) and Liouville ($\phi_0$) bosonic
zero modes we have\ref{2,3}
$$ {\cal A}_{\cal N}^{(n,{\cal N}-n)}=\Gamma (-s)\left(
{i\mu\over \pi}\right)^s\left\langle
\prod_{i=1}^n V_{-{1\over 2}}(k_i,\epsilon_i ) \prod_{j=n+1}^{\cal N}
T(k_j)\left(\int d^2\theta d^2z e^\Phi\right)^s\right\rangle_{\mu =0}
,\eqno(12)$$
where the momentum conservation law and the definition of $s$ are,
respectively,
$$\eqalign{\sum_ik_i&=2\alpha_0,\cr
\sum_i\beta_i&=-Q-\alpha_+s,\cr}\eqno(13)$$

As in the previous calculations\ref{8,9}, the strategy is to assume
that $s$ is a positive integer and to analytically continue the result for
arbitrary real $s$. For integer $s$, it is enough to use free
propagators to write (12) in an integral
representation. In the case of no spinor emission vertex $(n=0)$ these
amplitudes have been calculated in ref. [9] for arbitrary value of $\cal N$ by
successive derivatives of the ${\cal N}=3$ point function with respect to the
cosmological constant $\mu$, the ${\cal N}=3$ point integral was calculated
on its turn in ref.[7,8]. In the case where we include the spinor
emission vertex $V_{-1/2}^\epsilon$  we have not been able to accomplish the
integrals corresponding to the three point function\newfoot{$^{2)}$}{The only
non vanishing correlators are those with a net -2 charge for the $\sigma
$-ghost which cancels the background +2 charge.} $\langle V_{-1/2}^{\epsilon_1}
V_{-1/2}^{\epsilon_2}T^{-1}_k\rangle$ for  non-vanishing integer $s$ and
therefore we cannot use the same strategy as above to obtain higher point
amplitudes from the 3-point amplitude, hence we restrict ourselves in this
paper
to the so called bulk amplitudes $(s=0)$ where we take the finite part of
$\Gamma(-s)\mu^s$ when $s\to 0$, namely $\ln \mu$. The simplest non trivial
bulk amplitude to be calculated (see ref.[3]) is the two spinor-two scalar
scattering four-point amplitude:
$$
{\cal A}_4^{(2,2)} = \ln \mu \left\langle V_{-1/2}^{\epsilon _1}(k_1)V_{-1/2}^
{\epsilon _2}(k_2)T^{-1}_{k_3}T_{k_4}\right\rangle \quad .
\eqno(14)$$
It is convenient at this point to rewrite $T_{k_4}$   from (4) in the bosonized
form, using (8), to obtain
$$
T_{k_4}={1\over 2}\int d^2z_4 e^{ik_4x + \beta_4\phi}
\left(\alpha_+e^{ih(z_4)} +2\alpha_-m_4e^{-ih(z_4)}\right)
\left(\alpha_+e^{ih(\bar z_4)} +2\alpha_-m_4e^{-ih(\bar z_4)}\right)
\eqno(15)$$
where $m_i={1\over 2}(\beta_i^2-k_i^2)$. From the conservation law derived
from the integration of the zero mode associated with the massless field $h$ we
must have $\epsilon _1=\epsilon _2$ in order to have ${\cal A}_4^{(2,2)}\ne
0$; we
choose for convenience $\epsilon _1=\epsilon _2= + 1$. In this case if we
follow Seiberg\ref{16}
and work only with positive energy particles $(E>0)$ we use (11) to obtain
$k_1,k_2\ge \alpha_0$. The remaining two momenta $k_3$ and $k_4$
are chosen such that $k_3\ge \alpha_0$ and $k_4\le \alpha_0$, thus after fixing
the residual $\widehat{SL}(2,\Re )$
invariance with $z_1=0, z_2=1, z_3=\infty $ and calling $z_4=z$ we have
$$
\eqalign{
{\cal A}^{(2,2)}_4=&
2\ln\mu \;\alpha_-^2m_4^2 \int d^2z \vert z \vert ^{2\theta_{14}-1}
\vert 1-z\vert ^{2\theta_{24}-1}\cr
=& 2\pi\ln\mu\,\alpha^2_-m_4^2 \Delta (\theta_{14}+{1\over 2})\Delta
(\theta_{24}
+{1\over 2}) \Delta (-\theta_{14}- \theta_{24})\cr}
\eqno(16)$$
where $\theta_{ij}=k_ik_j-\beta_i\beta_j$ and $ \Delta (x)
=\Gamma(x)/\Gamma(1-x)$. Using  (13)
with $s=0$, we can rewrite ${\cal A}_4^{(2,2)}$ in the
kinematic region $k_i\ge \alpha_0\, ,\, i=1,2,3\, ,\, k_4\le \alpha_0$:
$$
{\cal A}_{4}^{(2,2)}= 2\pi\ln\mu\alpha^2_-\Delta (m_1)\Delta(m_2)
\Delta(m_3+1/2)\quad .\eqno(17)$$
This result agrees with refs.[3] and suggests again the
redefinition\newfoot{\ref{3)}}{Note that due to the kinematics, $m_4=-{1\over
2}$, and the factor $\Delta(0)$ we would be dividing by corresponds to $\Gamma
(-s=0)$, which has been combined with $\mu^s$ in order to produce $\ln\mu$.}:
$$
\eqalign{
T_k&\to T_k/\Delta(m+1/2)\cr
V^\epsilon _{-1/2}(k)&\to V^\epsilon _{-1/2}(k)/\Delta(m)\quad .\cr}\eqno(18)$$
This is in agreement (up to a factor $\pi \alpha_-^2$)
with the conjectured result (for ${\cal N}=4$ and $s=0$):
$$
{\cal A}_{\cal N}^{(n,{\cal N}-n)} =
{\partial ^{{\cal N}-3}\over \partial \mu^{{\cal N}-3}}
\mu ^{s+{\cal N}-3}\eqno(19)
$$

The next simplest amplitude to calculate is $\left\langle V_{\epsilon
_1}^{-1/2}
V_{\epsilon _2}^{-1/2}V_{\epsilon _3}^{-1/2}V_{\epsilon
_4}^{-1/2}\right\rangle$
which
cannot be calculated directly since the corresponding integral does not
converge in the kinematic region $k_1,k_2\ge \alpha_0$ $k_3,k_1 \le \alpha_0$
which is required by selection rule imposed by the zero mode of $h$.
Therefore, we treat instead the case of the 4 spinor - 1 scalar
scattering $({\cal A}_5^{(4,1)})$,
$${\cal A}_5^{(4,1)}=\left\langle
V^{\epsilon_1}_{-{1\over 2}} (k_1)\cdots
V^{\epsilon_4}_{-{1\over 2}}(k_4)T_{k_5}\right\rangle$$
Due to the $h$ zero mode selection rule we need three spinors with the same
polarization and the last one with opposite
polarization to those three; thus we choose
$\epsilon_1=+1=-\epsilon_2=-\epsilon_3=-\epsilon_4$, and the kinematic region
$k_i\ge \alpha_0$, $(i=1,2,3,4)$, $k_5\le \alpha_0$. Fixing
the gauge $z_2=1$, $ z_4=\infty $,  $z_5=0$ and defining
$z_1= w $, $z_3=z$  we have:
$$\displaylines{{\cal A}_5^{(4,1)}=\ln \mu \;\alpha_+^2\int
d^2zd^2w\times\hfill\cr
\hfill\times\vert z\vert
^{2(2m_3-1)}\vert w\vert ^{2(2m_1)}\vert 1-z\vert ^{2(-m_2-m_3)}\vert 1-w\vert
^{2(-m_2-m_1-1/2)}\vert z-w\vert ^{2(-m_3-m_1-1/2)}\quad (20)\cr}$$

In general, we do not know how to calculate such integrals, but in this
specific case we find that after a convenient shift in $m_1$ the above integral
can be cast into the form of the bulk five point correlator  of scalars
in the bosonic non-critical string, and such integral has been indirectly
calculated in [3]; using that result, we arrive at
$${\cal A}_5^{(4,1)} =\ln\mu\alpha_+^2\Delta (m_1+{1\over
2})\prod_{i=2}^4\Delta(m_i)\; .\eqno(21)$$
The above result also confirms formula (19) (up to $\alpha_+^2$) after the
redefinitions (18).

Now we come to the main computation of this paper, namely the two spinors-three
scalars scattering amplitude:
$${\cal A}_5^{(2,3)}=\left\langle
V^{\epsilon_1}_{-{1\over 2}}
 V^{\epsilon_2}_{-{1\over 2}}T_{k_3}^{(-1)}T_{k_4}
 T_{k_5}\right\rangle\eqno (22)$$

Due to the $h$ zero mode, the spinors must have opposite helicity,
and we choose $\epsilon_1=+1=-\epsilon_2$, and the kinematic region
$k_1\le\alpha_0$, $k_i\ge\alpha_0$, $i=2,3,4,5$. Fixing the gauge $z_1=0$,
$z_2=\infty$ and $z_3=1$, and defining $z_4=z$, $z_5=w$, it is easy
to arrive at the following integral representation :

$${\cal A}_5^{(2,3)}
=\int d^2 w \int d^2 z \vert z\vert^{2(2m_4-{1\over
2})} \vert w\vert^{2(2m_5-{1\over 2})}  \vert 1-z\vert^{2(-m_3-m_4)} \vert
1-w\vert^{2(-m_3-m_5)} \times\vartheta .\eqno(23)$$
Where the fermionic correlators ($\vartheta$) are given by the expression
$$\displaylines{\vartheta ={1\over 4} \left\langle
e^{{i\over 2}h(0)} e^{-{i\over 2}h(\infty )}
\left(\beta_4\psi (z) +ik_4\xi (z)\right)\left(\beta_5
\psi (w) +ik_5\xi (w)\right) \right\rangle \times
 \Bigl[ h.c.\Bigr] =\hfill\cr
 = {1\over 4}\Biggl\langle e^{{i\over 2}h(0)} e^{-{i\over 2}h(\infty )}
\left( e^{ih(z)}\mathop{\underbrace{(\beta_4-k_4)}}_{\alpha_+}
+ e^{-ih(z)}\mathop{\underbrace{(\beta_4+k_4)}}_{2\alpha_-m_4}
\right)\times\hfill\cr
\hfill\left( e^{ih(w)}\mathop{\underbrace{(\beta_5-k_5)}}_{\alpha_+}
+ e^{-ih(w)}\mathop{\underbrace{(\beta_5+k_5)}}_{2\alpha_-m_5} \right)
\Biggr\rangle \times \Bigl[ h.c.\Bigr] \cr
\hfill = \left[\left(m_4+m_5\right)\left( m_4\vert z\vert^{-1}\vert w\vert
+m_5\vert z\vert\vert w\vert^{-1}\right)\vert z-w\vert^{-2} -m_4m_5\vert
z\vert^{-1}\vert w\vert^{-1}\right] \quad (24)\cr}$$
Therefore we have,

$${\cal A}_5^{(2,3)}=
\left\{\left[ m_4(m_4+m_5)I_1 (m_3,m_4,m_5) +m_5(m_4+m_5)
I_1(m_3,m_5,m_4)\right] -m_4m_5I_2\right\} \quad \eqno (25)$$
where the first integral is given by the expression
$$I_1=\int d^2z\int d^2w \vert z\vert^{2(2m_4-1)} \vert w\vert^{2(2m_5)}
\vert 1-z\vert^{2(-m_3-m_4)} \vert 1-w\vert^{2(-m_3-m_5)} \vert z-w\vert^{2(
-m_4-m_5-1)}\eqno(26)$$
while the second one has already appeared in bosonic computations\ref{3}:
$$I_2=\int d^2z\int d^2w \vert z\vert^{2(2m_4-1)}
\vert w\vert^{2(2m_5-1)} \vert 1-z\vert^{2(-m_3-m_4)}
\vert 1-w\vert^{2(-m_3-m_5)} \vert z-w\vert^{2(-m_4-m_5)}\eqno(27)$$
In general the amplitude ${\cal A}_5^{2,3}$ is a combination of two unknown
integrals, however it is very fortunate that in the case where we place one of
spinor vertices at the infinity we end up with the two integrals above where
$I_2$ has been calculated (indirectly) before in ref.[3] with the result:
$$
I_2=\pi^2\Delta(m_3)\Delta(m_4)\Delta(m_5)\Delta(1-m_3-m_4-m_5)=\pi^2
\Delta(m_3)\Delta(m_4)\Delta(m_5)\Delta(1/2+m_2)\eqno(28)
$$
where we have used the kinematic relation $m_2+m_3+m_4+m_5=1/2$. At this point
it is quite surprising to have the result (28) contributing
to ${\cal A}^{(2,3)}_5$ since the role of
the factors $\Delta(m)$ and $\Delta(m+1/2)$ seem to be interchanged which might
point to some non universality of the vertices redefinition (18).
We will convince the
reader by calculating $I_1$ that this is actually not the case and the
misplaced poles of $I_2$ are cancelled by corresponding
poles of $I_1$. In order to
obtain $I_1$ we follow a by now standard procedure. First of all it is
convenient to rewrite $I_1$ using a change of variables after which we have:
$$\displaylines{
I_1(m_3,m_4,m_5)=\int d^2z d^2w\times\hfill \cr
\hfill\times\vert z\vert ^{2(2m_3-1)}\vert w\vert
^{2(-m_3-m_5)}\vert 1-z\vert ^{2(2m_4-1)}\vert 1-w\vert ^{2(-m_4-m_5-1)}\vert
z-w\vert ^{2(2m_5)}\quad (29) \cr}
$$
Using a technique developed by Dotsenko and Fatteev\ref{21}
it is easy to derive from (26) and (29) the asymptotic behaviour:
$$\eqalign{I_1(m_3,m_4\to\infty ,m_5)&\sim m_4^{-2-2(m_3+m_5)}\cr
I_1(m_3,m_4,m_5\to\infty )&\sim m_5^{-2(m_3+m_4)}\cr
I_1(m_3\to\infty ,m_4,m_5)&\sim m_3^{-2(m_4+m_5)}\cr }\eqno(30)$$

Now we use the property that at the poles of intermediate states ${\cal A}_5$
factorizes in four point functions, for which we can use the Virasoro-Shapiro
formula; in other words by using $\vert z\vert^{-2+\epsilon}=
{\pi\over\epsilon}\delta^{(2)}(z)$ we get  from (26) and (29):
$$
\eqalign{
I_1(m_4=\epsilon)&= {\pi^2\over
\epsilon}\Delta(m_5)\Delta(m_3)\Delta(1-m_3-m_5)\cr
I_1(m_3=\epsilon)&= {\pi^2m_5^2\over
\epsilon (m_4+m_5)^2}\Delta(m_4)\Delta(m_5)\Delta(1-m_4-m_5)\cr
I_1(m_5=-1/2+\epsilon)&= {-\pi^2\over
\epsilon(m_4-1/2)^2}\Delta(1-m_3-m_4)\Delta(m_3+1/2)\Delta(m_4+1/2)\cr }
\eqno(31)$$
Furthermore for $m_5=0$ the two integrals over $z$ and $w$ decouple,
and we have (up to possible factors 1/2)
$$
I_1(m_5=0)= {-\pi^2\over
m_4^2}\Delta(1/2-m_3-m_4)\Delta(m_3+1/2)\Delta(m_4+1/2)
\eqno(32)$$
Finally $I_1$ is also calculable at $m_4+m_5=\epsilon $ where it  contains a
double pole:
$$
I_1(m_4+m_5=\epsilon)=-{1\over\epsilon^2}\quad .
\eqno(33)$$
By taking  formulae (31-33) and the result (28) for $I_2$ the reader
can check that the
poles at $m_3$, $m_4$, $m_5\sim \epsilon $ and the double pole at
$m_4,m_5=\epsilon$
are canceled and do not appear in the final expression for ${\cal
A}_5^{(2,3)}$,
only the pole at $m_5=-1/2 +\epsilon$ survives which is in agreement with our
expectations. The reader may object at this point that we have only checked the
calculation of the first poles of the gamma functions of $I_2$, however by
using
the factorization properties, i.e., $\vert z\vert ^{-6+\epsilon}\sim
{\pi\over\epsilon}(\partial _z\partial _{\overline z})^2\delta^{(2)}(z)$ we
have also obtained after a long algebra:
$$
I_1(m_3=-1+\epsilon)\approx -{m_5^2\over 6}(m_4+m_5-1)^2\Delta(m_4)
\Delta(m_5)\Delta(-m_4-m_5)\quad .   \eqno(34)
$$
Introducing the result above and (28) altogether in (25)
one checks that the first
excited pole at $m_3=-1$ of $I_2$ does not appear in ${\cal A}_5$ and this is a
very striking cancelation which leads us to suggest the following result for
$I_1$:
$$
\eqalign{
I_1=& {\pi^2m_5\over
(m_4+m_5)^2}\Delta(m_3)\Delta(m_4)\Delta(m_5)\Delta(1-m_3-m_4-m_5)\cr
&- {\pi^2\over(m_4+m_5)^2}\Delta(m_3+1/2)\Delta(m_4+1/2)\Delta(m_5+1/2)
\Delta(1/2-m_3-m_4-m_5)\cr}\eqno(35)
$$
The above result is in agreement with all\newfoot{$^{4)}$}{Formula (35) does
not
agree with $I_1(m_4+m_5=\epsilon)$ and $I_2(m_5=-1/2+\epsilon)$ up to factors
1/2 which we conjecture to  be traced back to
symmetry factors.} formulae we have derived so
far from $I_1$ including the asymptotic behaviors (30) (one uses Stirling
formula  to check  it). Using (28) and (35) in (25) we have as expected:
$$ {\cal A}_5^{(2,3)}=
-\pi^2\Delta(m_2)\Delta(m_3+1/2)\Delta(m_4+1/2)\Delta(m_5+1/2)\quad.\eqno(36)$$

At last we should remark that although we have not been able to explicitly
calculate the four spinor bulk $(s=0)$ scattering ${\cal A}_4^{(4,0)}$ it is
clear that for $s=1$ it corresponds to ${\cal A}_5^{(4,1)}$ with $k_1=0$. In
order to calculate correlators involving more than four spinors we need the
$V_{+1/2}$ vertex but this is out of the scope of this letter.

{}From the result (36) we are reasonably safe to conclude that the correlation
functions factorize, and each bosonic external leg contributes a factor $\Delta
(m+{1\over 2})$, while a Ramond vertex gives a contribution $\Delta (m)$.

The factorizable result is actually highly non-trivial. Although, as mentioned,
some authors already present some of them, it is not clear at all how they
should be obtained from the recent supermatrix model techniques\ref{11}, in
such a way that a non-trivial check is also required. In some cases, the
relevant spinor fields seem to contribute with a zero factor to several
correlators (those involving more than two spinors) as opposite to our
non-vanishing result (21), which involves four spinors.
Nevertheless, there is a possibility of matching the results
disentangling the analytical contributions to the fermionic
correlators in a similar way as it was done in the bosonic matrix
models\ref{12}.
\vskip1cm

\centerline{\bf References.}
\vskip.5cm
\refer[[1]/D. J. Gross, A. A. Migdal {\it Phys. Rev. Lett.} {\bf 64} (1990)127;
E. Br\'ezin, V. A. Kazakov {\it Phys. Lett. } {\bf 236B} (1990)144; M.D.
Douglas and S.H. Shenker {\it Nucl. Phys.} {\bf B335} (1990)635.]
\refer[[2]/M. Goulian, M. Li {\it Phys. Rev. Lett.} {\bf 66} (1991)2051; Vl. S.
Dotsenko, {\it Mod. Phys. Lett.} {\bf A6} (1991)3601; K. Aoki, E. D'Hoker
{\it Mod. Phys. Lett. } {\bf A7} (1992)235.]
\refer[[3]/P. di Francesco, D. Kutasov { \it Nucl. Phys.} {\bf B375}
(1992)119.]
\refer[[4]/ E. Witten, {\it Nucl. Phys. } {\bf B373} (1992)187; M. Bershadsky,
talk given at Summer School of High Energy Physics and Cosmology, Trieste,
June/1992 (ICTP-prep); S. Govindarajan, J. Jayaraman and V. John Madras prep
IMSc-92/30, hepth@xxx/ 9207109.]
\refer[[5]/J.L. Gervais Paris prep-LPTENS-91/22, hepth@xxx/9205034.]
\refer[[6]/I. Klebanov, Lecture at the ICTP Spring School on String Theory
and Quantum Gravity, Trieste, April 1991, and princeton prep PUPT-1271, July
1991.]
\refer[[7]/L. Alvarez Gaum\`e and Ph. Zhaugg, {\it Phys. Lett.} {\bf B273}
(1991)81; K. Aoki, E. D'Hoker {\it Mod. Phys. Lett. } {\bf A7} (1992)333.]
\refer[[8]/E. Abdalla, M. C. B. Abdalla, D. Dalmazi, K. Harada
{\it Phys. Rev. Lett.} {\bf 68} (1992)1641.]
\refer[[9]/E. Abdalla, M. C. B. Abdalla, D. Dalmazi, K. Harada
{Int. J. Mod. Phys.} {\bf A7} (1992)7339.]
\refer[[10]/E. Abdalla, M.C.B. Abdalla and D. Dalmazi {\it Phys. Lett. }
{\bf 291B} (1992)38; D. Dalmazi in summer workshop in high energy physics and
cosmology, Trieste, 1992.]
\refer[[11]/L. Alvarez Gaum\'e, H. Itoyama, J. L. Ma\~nez, A. Zadra {Int. J.
Mod. Phys.} {\bf A7} (1992)5337; L. Alvarez Gaum\'e, K. Becker, M. Becker, R.
Emparan and J. Ma\~nes, CERN prep TH-6687/92]
\refer[[12]/G. Moore, N. Seiberg, M. Staudacher
{ \it Nucl. Phys.} {\bf B362} (1991)665.]
\refer[[13]/E. Abdalla, M.C.B. Abdalla, D. Dalmazi, A. Zadra, in preparation.]
\refer[[14]/J. Distler, Z. Lhousek, H. Kawai {\it Int. J. Mod. Phys.} {\bf A5}
(1990)391.]
\refer[[15]/D. Friedan, E. Martinec, S. Shenker {\it Nucl. Phys. } {\bf B271}
(1986)93; G. T. Horowitz, S. P. Martin, R. C. Myers {\it Phys. Lett.} {\bf
B215} (1988)291; {\bf B218} (1989)309.]
\refer[[16]/N. Seiberg, {\it Progr. Theor. Phys. Supp.} {\bf 102} (1990)319.]

\end